\documentclass[conference]{IEEEtran}
\IEEEoverridecommandlockouts
\usepackage{cite}
\usepackage{amsmath,amssymb,amsfonts}
\usepackage{algorithmic}
\usepackage{graphicx}
\usepackage{textcomp}
\usepackage{xcolor}

\usepackage{algorithm}
\usepackage{bm}
\usepackage{multirow}
\usepackage{booktabs}
\usepackage{enumitem}
\usepackage{url}

\def\BibTeX{{\rm B\kern-.05em{\sc i\kern-.025em b}\kern-.08em
    T\kern-.1667em\lower.7ex\hbox{E}\kern-.125emX}}
\begin{document}

\title{Towards Clinical Practice in CT-Based Pulmonary Disease Screening:\\An Efficient and Reliable Framework}


\author{\IEEEauthorblockN{
Qian Shao$^{*,1}$,
Bang Du$^{*,1}$,
Yixuan Wu$^{1}$,
Zepeng Li$^{1}$,
Qiyuan Chen$^{1}$,
Qianqian Tang$^{2}$,\\
Jian Wu$^{1}$,
Jintai Chen$^{\dagger,3}$,
Hongxia Xu$^{\dagger,1}$
\thanks{$^*$Equal contributions. $^{\dagger}$Corresponding author.} 
}
\IEEEauthorblockA{
\text{$^{1}$Zhejiang University}
\text{$^{2}$Shaoxing Tangtang Technology Co., Ltd.}
\text{$^{3}$The Hong Kong University of Science and Technology}
\\
\text{Email:\{qianshao, wujian2000, einstein\}@zju.edu.cn}  \\
}
}

\maketitle

\begin{abstract}
Deep learning models for pulmonary disease screening from Computed Tomography (CT) scans promise to alleviate the immense workload on radiologists. Still, their high computational cost, stemming from processing entire 3D volumes, remains a major barrier to widespread clinical adoption. Current sub-sampling techniques often compromise diagnostic integrity by introducing artifacts or discarding critical information. To overcome these limitations, we propose an Efficient and Reliable Framework (ERF) that fundamentally improves the practicality of automated CT analysis. Our framework introduces two core innovations: (1) A Cluster-based Sub-Sampling (CSS) method that efficiently selects a compact yet comprehensive subset of CT slices by optimizing for both representativeness and diversity. By integrating an efficient $k$-nearest neighbor search with an iterative refinement process, CSS bypasses the computational bottlenecks of previous methods while preserving vital diagnostic features. (2) An Ambiguity-aware Uncertainty Quantification (AUQ) mechanism, which enhances reliability by specifically targeting data ambiguity arising from subtle lesions and artifacts. Unlike standard uncertainty measures, AUQ leverages the predictive discrepancy between auxiliary classifiers to construct a specialized ambiguity score. By maximizing this discrepancy during training, the system effectively flags ambiguous samples where the model lacks confidence due to visual noise or intricate pathologies. Validated on two public datasets with $2,654$ CT volumes across diagnostic tasks for $3$ pulmonary diseases, ERF achieves diagnostic performance comparable to the full-volume analysis (over $90\%$ accuracy and recall) while reducing processing time by more than $60\%$. This work represents a significant step towards deploying fast, accurate, and trustworthy AI-powered screening tools in time-sensitive clinical settings.
\end{abstract}

\begin{IEEEkeywords}
Pulmonary disease screening, sub-sampling, uncertainty quantification
\end{IEEEkeywords}

\section{Introduction}

\label{sec:intro}

Deep learning has shown great promise in automating pulmonary disease screening from CT scans, potentially alleviating the immense workload on radiologists~\cite{hao2025local,tawfeek2025enhancing}.
However, the standard practice of processing entire 3D volumes is computationally prohibitive, hindering deployment in time-sensitive or resource-constrained clinical settings~\cite{balasamy2025hco,pham2025leveraging}.
Unlike radiologists, who quickly skim to focus on relevant slices, full-volume approaches waste resources on diagnostically irrelevant data.
Therefore, developing intelligent methods to select salient diagnostic information is crucial for practical clinical integration.

\begin{figure}[tbp]
    \includegraphics[width=1\linewidth]{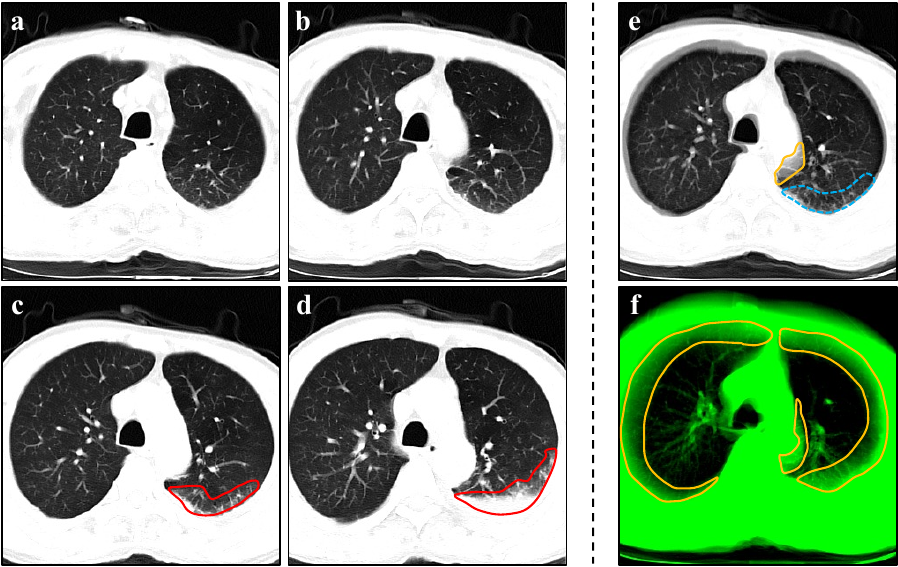}
    \caption{Visualization of interpolation methods on common pneumonia (CP) lesions. (\textbf{a-d}) are four consecutive slices showing CP lesions (red solid-line area). (\textbf{e}) and (\textbf{f}) are results of spline interpolation and projection interpolation, respectively. Yellow solid lines indicate processing-induced artifacts, blue dashed lines highlight regions of lesion elimination post-processing.}
    \label{f1}
    \vspace{-3mm}
\end{figure}

Prior work on sub-sampling often compromises diagnostic integrity.
Interpolation techniques can introduce artifacts or erase critical details~\cite{liauchuk2019imageclef}, as shown in Fig.~\ref{f1}, while conventional slice selection relies on rigid heuristics~\cite{zunair2019estimating,zunair2020uniformizing}.
Although recent machine learning utilizes representativeness or diversity~\cite{sener2018active,wu2024optimal}, they face three key limitations:
(1) Focusing exclusively on representativeness causes redundancy, while solely prioritizing diversity risks missing collective diagnostic signals~\cite{zhang2023model}.
(2) Calculating pairwise similarities for selection contradicts the goal of efficiency~\cite{xie2023active,wang2022unsupervised}.
(3) Neglecting lung anatomy leads to incoherent sub-sampling, degrading downstream performance~\cite{zunair2020uniformizing}.

The challenge of using reduced data subsets elevates the need for robust uncertainty quantification (UQ).
However, existing UQ methods like Bayesian Neural Networks~\cite{ngartera2024application} or Deep Ensembles (DE)~\cite{lakshminarayanan2017simple} are computationally too expensive for efficient screening pipelines~\cite{abboud2024sparse}.
More importantly, standard metrics like predictive entropy often fail to specifically capture data ambiguity arising from subtle lesions or artifacts, which are common in medical imaging.
Without explicitly flagging these ambiguous cases, a model operating on partial data risks making overconfident yet incorrect predictions.

To address these challenges, we propose an Efficient and Reliable Framework (ERF).
First, to ensure \textbf{diagnostic integrity efficiently}, we design a Cluster-based Sub-Sampling (CSS) method.
CSS first partitions the lung volume based on anatomical structure.
Within each part, CSS employs $K$-Means to group slices~\cite{wang2022unsupervised} and an efficient approximate $k$-NN search algorithm based on Hierarchical Navigable Small World (HNSW) graphs~\cite{malkov2018efficient} to identify the most representative slice in each cluster.
An iterative refinement process then optimizes for diversity, ensuring the subset is both compact and comprehensive without heavy computation.
Second, to guarantee \textbf{reliability}, we introduce an Ambiguity-aware Uncertainty Quantification (AUQ) mechanism.
Inspired by discrepancy-maximization in domain adaptation~\cite{saito2018maximum,cho2022mcdal}, we employ two auxiliary classifiers alongside the primary model.
By training these auxiliary heads to maximize their predictive discrepancy, we force them to expose decision boundaries sensitive to ambiguity.
This discrepancy serves as a precise proxy for data ambiguity, which we combine with the primary classifier's predictive entropy to form a robust, low-overhead uncertainty score.

We validate ERF on two public datasets totaling $2,654$ CT volumes for three pulmonary diseases.
Experiments demonstrate that ERF achieves diagnostic performance (over $90\%$ accuracy and recall) comparable to full-volume analysis while reducing processing time by over $60\%$.
Furthermore, the visualization results demonstrate that CSS can select representative CT slices containing crucial indicators to aid in diagnosis.
The main contributions are:
\begin{itemize}[leftmargin=*]
  \item We develop ERF, a framework balancing high efficiency with diagnostic accuracy for pulmonary screening.
  \item We propose CSS to hierarchically select representative and diverse CT slices, bypassing computational bottlenecks of prior methods.
  \item We design AUQ, which leverages classifier discrepancy to specifically quantify data ambiguity and model uncertainty with minimal overhead.
  \item Extensive validation confirms ERF matches full-volume performance with significantly reduced computational cost.
\end{itemize}
\section{Related Works}
\label{sec:rw}

\subsection{Pulmonary Disease Screening}
Deep learning has established 3D CT as the preferred modality for pulmonary screening due to its rich anatomical detail~\cite{fink2025artificial,tawfeek2025enhancing}.
State-of-the-art models have achieved remarkable success in detecting diverse pathologies, ranging from interstitial and obstructive diseases~\cite{guiot2025automated,zhu2024advancements} to neoplastic and infectious conditions~\cite{quanyang2024artificial,kordnoori2025lungxpertai}.
However, the substantial computational overhead of processing high-resolution volumetric data remains a critical bottleneck.

\subsection{Sub-Sampling}

To alleviate the computational burden, data reduction techniques are widely adopted. 
Classical approaches rely on interpolation~\cite{needham1959science,de1962bicubic,liauchuk2019imageclef} or heuristic slice selection~\cite{zunair2019estimating,zunair2020uniformizing}. 
While interpolation acts as a generalized sub-sampling method, it risks introducing artifacts, whereas heuristic selection often disrupts volumetric continuity, potentially omitting small but critical lesions.

Learning-based sampling offers a more sophisticated alternative, generally categorized by objectives of representativeness or diversity. 
Representative sampling minimizes statistical discrepancies between the subset and the full volume, \textit{e.g.}, Wasserstein~\cite{graf2007foundations} or energy~\cite{mak2018support} distance, maximum mean discrepancy~\cite{chen2012super}, and generalized empirical $F$-discrepancy~\cite{zhang2023model}.
Conversely, diversity sampling seeks to maximize information content by selecting varied samples~\cite{wu2024optimal}. 
Despite these advancements, generic sampling algorithms typically disregard task-specific anatomical priors. They treat CT scans as unstructured data distributions, failing to preserve the structural context essential for accurate medical diagnosis.

\subsection{Uncertainty Quantification}

UQ~\cite{hullermeier2021aleatoric,wang2025aleatoric} is vital for deploying deep learning models in critical applications like medical diagnosis, enabling them to signal low confidence.
Prominent UQ approaches include Bayesian Neural Networks~\cite{neal2012bayesian,ngartera2024application}, Monte Carlo (MC) Dropout~\cite{gal2016dropout}, and DE~\cite{lakshminarayanan2017simple}. While powerful, these methods introduce significant computational overhead, such as requiring multiple forward passes for inference or training multiple models, which dramatically increases resource consumption~\cite{abdar2021review}.
Other notable methods include evidential deep learning~\cite{sensoy2018evidential}, which aims for single-pass uncertainty, and conformal prediction~\cite{vovk2005algorithmic}, providing model-agnostic guarantees but whose utility depends on calibration.
\section{Methods}
\label{method}

Our proposed ERF operates in two stages: CSS first compresses volumetric data into informative slices via iterative density-diversity optimization. Subsequently, AUQ employs a discrepancy-maximization mechanism to perform robust classification while explicitly flagging data ambiguity.

\subsection{Cluster-based Sub-Sampling}
\label{css}
Given a dataset of $N$ CT volumes $\mathcal{D}=\{(\bm{x}_i,\bm{y}_i)\}_{i=1}^N$, where each volume $\bm{x}_i$ contains $n$ slices, our goal is to select a subset of $m \ll n$ representative and diverse slices to approximate the full-volume diagnostic performance. CSS achieves this through a structured, iterative process involving anatomical partitioning, clustering, and refined selection.

\subsubsection{Partitioning and Representation}
To preserve anatomical semantics, we first partition each volume into upper, middle, and lower regions based on a fixed ratio ($0.25:0.15:0.6$), distributing the selection budget $m$ proportionally.
Subsequently, we map each slice $\bm{s}_j$ to a feature vector $\bm{h}_j \in \mathbb{R}^d$ using the image encoder of a frozen, pre-trained MedCLIP-ViT~\cite{wang2022medclip}, normalized such that $\Vert \bm{h}_j\Vert_2=1$. This inference-only step ensures high-level semantic capture with minimal computational overhead.

\subsubsection{Iterative Selection via Density and Diversity}
CSS operates on the principle that selected slices should be both representative (density peaks) and diverse (covering the feature space).
First, we partition the slice features within each anatomical region into clusters using $K$-Means. Let $C_k$ denote the $k$-th cluster. We aim to select one instance $\bm{z} \in C_k$ that best balances local density and global diversity.

\paragraph{Representativeness}
We estimate the local density of an instance $\bm{z}$ using the Euclidean distance to its $k$-NN, denoted as $D(\bm{z}, \bm{z}|k)$. Minimizing this distance identifies density peaks. To ensure scalability, we employ Hierarchical Navigable Small World (HNSW) graphs~\cite{malkov2018efficient} for approximate $k$-NN search.

\paragraph{Diversity Refinement}

Selecting only density peaks may yield redundant slices if clusters are semantically close. We introduce an iterative refinement process. Let $Z^{(t)}$ be the set of selected instances at iteration $t$. For a candidate instance $\bm{z}$ in cluster $C_k$, we define a diversity regularizer $\Phi(\bm{z}, t)$ that penalizes proximity to instances selected from \textit{other} clusters in the previous iteration:
\begin{equation}
\Phi(\bm{z}, t) = \sum\limits_{\tilde{\bm{z}} \in Z^{(t-1)} \setminus C_k} \frac{1}{\Vert \bm{z} - \tilde{\bm{z}} \Vert_2^\alpha},
\end{equation}
where $Z^{(t-1)} \setminus C_k$ denotes the set of previously selected instances excluding the one belonging to cluster $C_k$, and $\alpha$ is a sensitivity hyperparameter controlling the strength of the distance penalty.
This term is stabilized via an Exponential Moving Average (EMA)~\cite{laine2016temporal} with momentum $\beta$:
\begin{equation}
    \overline{\Phi}(\bm{z}, t) = \beta \cdot \overline{\Phi} (\bm{z}, t-1) + (1-\beta) \cdot \Phi(\bm{z}, t).
\end{equation}

\paragraph{Selection Objective}
At each iteration $t$, we update the selection for cluster $C_k$ by maximizing a joint objective:
\begin{equation}
\label{eq:joint_obj}
\widetilde{\bm{z}}^{(k,t)} = \mathop{\arg\max}\limits_{\bm{z} \in \mathcal{N}_h(\widetilde{\bm{z}}^{(k,t-1)})} \left[ \frac{1}{D(\bm{z}, \bm{z}|k)} - \lambda \cdot \overline{\Phi}(\bm{z}, t) \right],
\end{equation}
where $\lambda$ balances representativeness and diversity. To further accelerate computation, the search space is restricted to the $h$ nearest neighbors $\mathcal{N}_h$ of the currently selected instance. The final subset $\widetilde{\mathcal{D}}$ is formed after $T$ iterations.

\begin{figure}[tbp]
  \includegraphics[width=\linewidth]{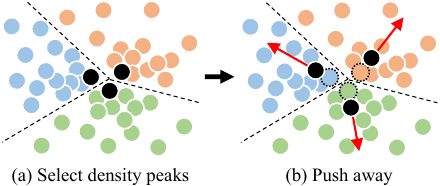}
  \caption{The core insight of CSS. Density peaks are selected for representativeness, and then they are pushed away via an iterative refinement process for diversity. The dashed and solid black circles denote the selected instances from the previous iteration and the current iteration, respectively.}
  \label{fcss}
  \vspace{-3mm}
\end{figure}

\subsection{Ambiguity-aware Uncertainty Quantification (AUQ)}
\label{auq}
While the selected subset $\widetilde{\mathcal{D}}$ enables efficient screening, sub-sampling inherently introduces ambiguity. Standard entropy metrics often fail to capture ambiguity, \textit{e.g.}, artifacts, subtle lesions. We propose AUQ to explicitly quantify this ambiguity via classifier discrepancy.

\paragraph{Framework}
We augment the primary classifier $G$ with two auxiliary classifiers $G_1$ with parameter matrix $\bm{W}_1$ and $G_2$ with $\bm{W}_2$.
All classifiers share the feature extractor $F$ but possess distinct prediction heads. The primary model is trained via standard cross-entropy $\mathcal{L}_{CE}$. The model architecture is shown in the Technical Appendix.

\paragraph{Discrepancy Maximization}
To capture ambiguity, we force $G_1$ and $G_2$ to diverge on ambiguous samples while maintaining accuracy. We define the discrepancy $d_{dis}$ as the mean $L_1$-distance between the probability vectors $\bm{p}^1, \bm{p}^2$ and the primary prediction $\bm{p}$, following~\cite{ben2010theory}:
\begin{equation}
\label{ddis}
    d_{dis}(\bm{p}_i^1,\bm{p}_i^2)=\frac{1}{N_{\text{cls}}}\left(\Vert\bm{p}_i^1-\bm{p}_i\Vert_1+\Vert\bm{p}_i^2-\bm{p}_i\Vert_1+\Vert\bm{p}_i^1-\bm{p}_i^2\Vert_1\right),
\end{equation}
where $N_{\text{cls}}$ is the number of classes.
To operationalize this, we formulate the training objective for the auxiliary classifiers:
\begin{equation}
\min\limits_{\bm{W}_1,\bm{W}_2} \frac{1}{N}\sum\limits^N_{i=1} \left( \mathcal{L}_{CE}(\bm{p}_i^1,\bm{y}_i) + \mathcal{L}_{CE}(\bm{p}_i^2,\bm{y}_i) - d_{dis}(\bm{p}_i^1,\bm{p}_i^2) \right).
\end{equation}
Crucially, gradients from the discrepancy term are detached from the backbone $F$ to prevent degrading the shared representation.

\paragraph{Uncertainty Score}
High discrepancy indicates that the input $\widetilde{\bm{x}}\in\widetilde{\mathcal{D}}$ lies near decision boundaries sensitive to subtle variations. We combine this ambiguity proxy with predictive entropy $\mathcal{H}$ for a robust uncertainty score $U(\widetilde{\bm{x}})$:
\begin{equation}
    U(\widetilde{\bm{x}})=d_{dis}(\bm{p}^1,\bm{p}^2)+\mathcal{H}(\widetilde{\bm{x}}).
\end{equation}
This score allows ERF to flag uncertain cases, providing a more reliable basis for
clinical decision-making.

\begin{figure}[tbp]
    \includegraphics[width=1\linewidth]{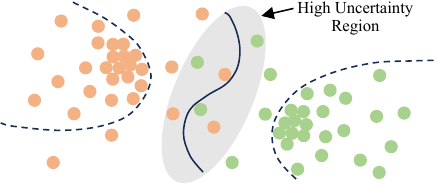}
    \caption{Illustration of AUQ. The circles of different colors represent samples from distinct categories. The solid and dashed lines denote the decision hyperplanes generated by the main and auxiliary classifiers, respectively. High-uncertainty samples are located in the gray area.}
    \label{fuq}
    \vspace{-3mm}
\end{figure}
\section{Experiments}

\begin{table*}[tbp]
\centering
\caption{Comparison with other sub-sampling methods. The best performance is bold, and the second-best performance is underlined, except for the Full CT method. Each result shows mean accuracy and standard deviation over $5$-fold cross-validation.}
\label{others}
\begin{tabular}{l|cc|cc|cc}
\toprule
\multicolumn{1}{c|}{}   & \multicolumn{2}{c|}{Task $1$}   & \multicolumn{2}{c|}{Task $2$}   & \multicolumn{2}{c}{Task $3$}   \\
\multicolumn{1}{l|}{\multirow{-2}{*}{Method}}
& Accuracy   & Recall   & Accuracy   & Recall   & Accuracy   & Recall   \\
\cmidrule{1-7}
Full CT
& $99.12_{\pm 0.31}$   & $99.15_{\pm 0.35}$
& $99.35_{\pm 0.41}$   & $99.30_{\pm 0.38}$
& $92.47_{\pm 0.60}$   & $92.10_{\pm 0.65}$   \\
\cmidrule{1-7}
PI
& $90.36_{\pm 0.34}$   & $89.02_{\pm 0.41}$
& $88.92_{\pm 0.40}$   & $89.00_{\pm 0.30}$
& $84.59_{\pm 0.47}$   & $83.50_{\pm 0.52}$   \\
LI
& $78.52_{\pm 2.54}$   & $77.89_{\pm 2.67}$
& $73.67_{\pm 2.03}$   & $71.70_{\pm 2.30}$
& $68.81_{\pm 3.80}$   & $67.54_{\pm 4.81}$   \\
SI
& \underline{$98.78_{\pm 0.34}$}   & $98.20_{\pm 0.37}$
& $98.33_{\pm 0.42}$   & $98.30_{\pm 0.40}$
& $87.24_{\pm 0.40}$   & $86.94_{\pm 0.37}$   \\
\cmidrule{1-7}
SSS
& $96.89_{\pm 0.56}$   & $96.28_{\pm 0.62}$
& $95.23_{\pm 0.97}$   & $95.12_{\pm 0.99}$            
& $89.34_{\pm 0.90}$   & $89.67_{\pm 1.24}$   \\
ESS
& $96.42_{\pm 0.86}$   & $96.85_{\pm 0.78}$
& $94.29_{\pm 1.37}$   & $94.50_{\pm 1.70}$
& $88.56_{\pm 1.56}$   & $86.78_{\pm 1.65}$   \\
\cmidrule{1-7}
CoreSet
& $97.38_{\pm 0.47}$   & $96.39_{\pm 0.59}$
& $98.06_{\pm 0.47}$   & $97.57_{\pm 0.77}$            
& $88.91_{\pm 1.88}$   & $89.07_{\pm 2.06}$   \\
ActiveFT
& $98.75_{\pm 0.38}$   & \underline{$98.26_{\pm 0.76}$}
& \underline{$98.62_{\pm 0.36}$}   & \underline{$98.38_{\pm 0.61}$}
& \underline{$90.68_{\pm 0.71}$}   & \underline{$90.35_{\pm 1.14}$}   \\
\cmidrule{1-7}
CSS (Ours)
& $\mathbf{98.89_{\pm 0.17}}$   & $\mathbf{99.05_{\pm 0.18}}$
& $\mathbf{98.65_{\pm 0.23}}$   & $\mathbf{98.82_{\pm 0.29}}$
& $\mathbf{91.88_{\pm 0.43}}$   & $\mathbf{90.85_{\pm 0.54}}$   \\
\bottomrule
\end{tabular}
\end{table*}

\subsection{Datasets and Tasks.}
We validate our framework on two public datasets: SARS-CoV-2~\cite{zhang2020clinically} and LUNG-PET-CT-Dx~\cite{li2020large}. Excluding volumes with insufficient depth ($<64$ slices), we curate a dataset of $2,654$ CT scans, comprising $747$ Novel Coronavirus Pneumonia (NCP), $741$ Common Pneumonia (CP), $311$ Adenocarcinoma (AC), and $855$ normal cases. We formulate three binary classification tasks to evaluate diagnostic versatility: Task $1$ (NCP vs. Normal), Task $2$ (CP vs. Normal), and Task $3$ (AC vs. Normal).
For each task, we employ $5$-fold cross-validation, reporting mean accuracy and recall.
Implementation details are provided in the Technical Appendix.

\begin{table}[tbp]
\vspace{-3mm}
\centering
\caption{Comparison with optimal methods of other types under a lower sub-sampling rate.}
\label{ec}
\begin{tabular}{l|c|cc}
\toprule
Task
& Method       & \multicolumn{1}{c}{Accuracy}     & \multicolumn{1}{c}{Recall}   \\
\cmidrule{1-4}
\multirow{4}{*}{Task $1$}
& SI           & $97.99_{\pm 0.85}$               & \underline{$98.06_{\pm 0.54}$}   \\
& SSS          & $95.38_{\pm 0.46}$               & $94.78_{\pm 0.56}$               \\
& ActiveFT     & \underline{$98.00_{\pm 0.48}$}   & $97.32_{\pm 0.47}$               \\
& CSS (Ours)   & $\mathbf{98.13_{\pm 0.66}}$      & $\mathbf{98.13_{\pm 0.30}}$      \\
\cmidrule{1-4}
\multirow{4}{*}{Task $2$}
& SI           & $96.73_{\pm 0.95}$               & $97.10_{\pm 0.79}$               \\
& SSS          & $94.11_{\pm 0.90}$               & $94.06_{\pm 0.89}$               \\
& ActiveFT     & \underline{$97.18_{\pm 0.22}$}   & \underline{$97.17_{\pm 0.56}$}   \\
& CSS (Ours)   & $\mathbf{97.68_{\pm 0.36}}$      & $\mathbf{98.11_{\pm 0.88}}$      \\
\cmidrule{1-4}
\multirow{4}{*}{Task $3$}
& SI           & $83.63_{\pm 1.24}$               & $83.49_{\pm 1.33}$               \\
& SSS          & $76.36_{\pm 1.68}$               & $77.16_{\pm 1.86}$               \\
& ActiveFT     & \underline{$84.73_{\pm 1.15}$}   & \underline{$85.21_{\pm 1.29}$}   \\
& CSS (Ours)   & $\mathbf{90.03_{\pm 0.40}}$      & $\mathbf{90.35_{\pm 1.14}}$      \\
\bottomrule
\end{tabular}
\vspace{-3mm}
\end{table}

\subsection{Comparison with Other Sub-Sampling Methods}

We compare CSS against different sub-sampling baselines.
The baselines include three widely-used interpolation methods: Projection Interpolation (PI)~\cite{liauchuk2019imageclef}, Linear Interpolation (LI)~\cite{needham1959science}, and Spline Interpolation (SI)~\cite{de1962bicubic},
two heuristic slice selection methods: Subset Slice Selection (SSS)~\cite{zunair2019estimating} and Even Slice Selection (ESS)~\cite{zunair2020uniformizing}, two representative and diverse sub-sampling methods: CoreSet~\cite{sener2018active} and ActiveFT~\cite{xie2023active}. We also include the full CT method as an upper-bound reference.
For a fair comparison, all sub-sampling methods select $64$ slices per volume, except for the full CT method, since it utilizes the complete CT data.
The same classification backbone and training hyperparameters are used for all experiments.

\begin{figure*}[tbp]
\centering
  \includegraphics[width=\textwidth]{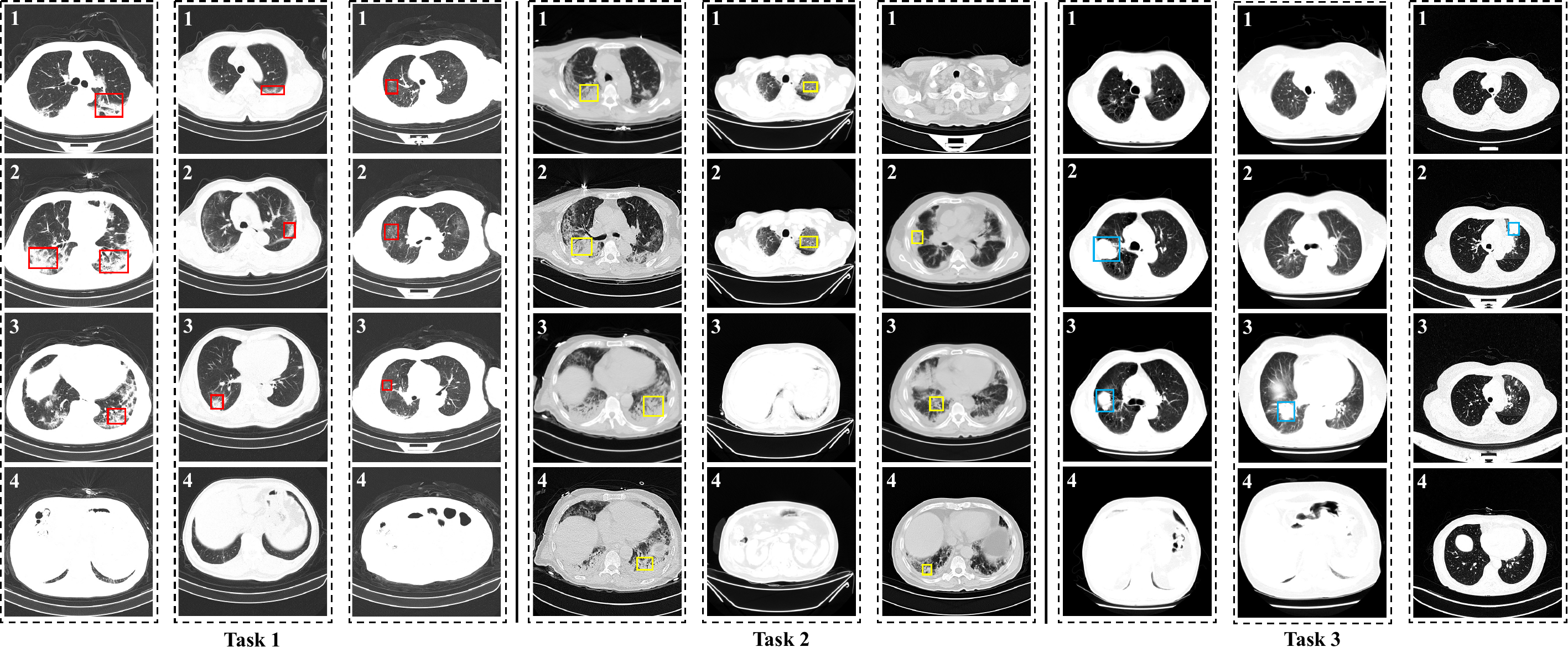}
  \caption{CT slices selected by CSS. The red, yellow, and blue boxes mark the lesion areas of NCP, CP, and AC, respectively.}
  \label{vis}
\end{figure*}

As shown in Table~\ref{others}, CSS consistently outperforms other sub-sampling methods, approaching the performance of using the full CT volume.
A notable performance drop is observed for all methods in Task $3$ (adenocarcinoma detection) compared to Tasks $1$ and $2$ (pneumonia diagnosis). We attribute this to the inherent difficulty of Task $3$, which involves identifying smaller, morphologically complex lesions, a challenge corroborated by our qualitative results in Fig.~\ref{vis}.

We further challenge our method by reducing the sampling budget to $32$ slices. 
As shown in Table~\ref{ec}, while other methods degrade rapidly, CSS exhibits strong robustness.
In Task $3$, CSS achieves a significant margin over the second-best ActiveFT ($+5.30\%$ accuracy, $+5.14\%$ recall).
This suggests that our strategy effectively identifies the most informative slices even when the budget is extremely tight. Additional experiments with $16, 8, 4$ slices are provided in the Appendix.

\begin{table}[tbp]
\vspace{-3mm}
\centering
\caption{Comparison with other UQ methods.}
\label{uq}
\begin{tabular}{l|ccc}
\toprule
Method         & Task $1$   & Task $2$   & Task $3$   \\
\cmidrule{1-4}
Entropy
& $50.79_{\pm 9.44}$                & $67.74_{\pm 10.78}$              & $57.95_{\pm 4.72}$   \\
Discrepancy
& \underline{$74.76_{\pm 15.91}$}   & $77.18_{\pm 10.25}$              & \underline{$77.69_{\pm 7.72}$}\\
DE
& $73.73_{\pm 16.41}$               & $69.60_{\pm 8.11}$               & $67.56_{\pm 8.89}$   \\
MC Dropout
& $70.95_{\pm 4.18}$                & \underline{$78.65_{\pm 5.88}$}   & $66.28_{\pm 8.07}$   \\
\cmidrule{1-4}
AUQ (Ours)
& $\mathbf{94.92_{\pm 7.05}}$       & $\mathbf{91.59_{\pm 7.93}}$      & $\mathbf{92.05_{\pm 5.45}}$   \\
\bottomrule
\end{tabular}
\vspace{-3mm}
\end{table}

\subsection{Comparison with Other UQ Methods}

We compare AUQ with Entropy, Discrepancy-only, Deep Ensemble (DE)~\cite{lakshminarayanan2017simple}, and MC Dropout~\cite{gal2016dropout}.
To quantify the effectiveness of UQ in flagging errors, we introduce misdiagnosis recall $\text{Recall}_{\text{mis}}$. 
Specifically, we select the top-$15$ most uncertain samples and calculate the recall of actual misclassified cases ($FN+FP$) within this subset. A higher $\text{Recall}_{\text{mis}}$ indicates the method correctly assigns high uncertainty to erroneous predictions.
To ensure fairness, we fix the backbone predictions across all UQ methods. 
Table~\ref{uq} demonstrates that AUQ significantly outperforms baselines, including computationally expensive DE, confirming that AUQ effectively captures data ambiguity.

\subsection{Efficiency Evaluation}

We analyze the computational complexity and actual inference time to demonstrate the efficiency of ERF.

\subsubsection{Complexity Analysis}
Table~\ref{tc} presents the asymptotic time complexity, where $M_{\text{DE}}$ represents the number of models that DE ensembles, and $M_{\text{MC}}$ represents the number of MC Dropout forward passes.
For sub-sampling, CSS ($O(N\cdot n \log n)$) is significantly more efficient than ActiveFT ($O(N \cdot n^2)$), as our HNSW-based search avoids exhaustive pairwise comparisons. 
For UQ, AUQ maintains the efficiency of single-pass methods, avoiding the linear cost scaling of DE ($M_{\text{DE}}$) and MC Dropout ($M_{\text{MC}}$).

\begin{table}[tbp]
\vspace{-3mm}
\centering
\caption{Time complexity comparison.}
\label{tc}
\begin{tabular}{l|c|l|c}
\toprule
Sub-Sampling & Complexity & UQ Method & Complexity \\
\cmidrule{1-4}
Full CT    & $O(N\cdot n)$        & Entropy     & $O(N\cdot n)$ \\
SI         & $O(N\cdot n)$        & Discrepancy & $O(N\cdot n)$   \\
SSS        & $O(N\cdot m)$        & DE          & $O(N\cdot n\cdot M_{\text{DE}})$ \\
ActiveFT   & $O(N\cdot n^2)$      & MC Dropout  & $O(N\cdot n\cdot M_{\text{MC}})$ \\
\cmidrule{1-4}
CSS (Ours) & $O(N\cdot n \log n)$ & AUQ (Ours)  & $O(N\cdot n)$ \\
\bottomrule
\end{tabular}
\end{table}

\begin{table}[tbp]
\vspace{-3mm}
\centering
\caption{Inference time for $50$ cases.}
\label{eff}
\begin{tabular}{l|ccc}
\toprule
Method       & Task $1$   & Task $2$   & Task $3$   \\
\cmidrule{1-4}
SI
& $52.83s \thinspace (31\%)$   & $55.35s \thinspace (29\%)$   & $53.68s \thinspace (30\%)$   \\
SSS
& $41.52s \thinspace (25\%)$   & $42.78s \thinspace (23\%)$   & $41.76s \thinspace (23\%)$   \\
ActiveFT
& $81.28s \thinspace (48\%)$   & $87.57s \thinspace (46\%)$   & $82.37s \thinspace (46\%)$   \\
CSS (Ours)
& $66.64s \thinspace (39\%)$   & $70.88s \thinspace (37\%)$   & $67.53s \thinspace (38\%)$   \\
\cmidrule{1-4}
Full CT
& $169.24s$ & $189.05s$ & $179.25s$   \\
\bottomrule
\end{tabular}
\vspace{-3mm}
\end{table}

\subsubsection{Runtime Comparison}

In addition, we calculate the inference times of CSS, and the optimal methods of other types (\textit{i.e.}, SI, SSS, and ActiveFT).
The inference time of ActiveFT and CSS comprises three components:
(1) Feature extraction time using the pre-trained MedCLIP-ViT;
(2) Time taken to select $32$ slices based on their respective strategies;
(3) Time for disease detection by the classification model.
The inference time of the SI and SSS methods only includes the time for slice selection (or interpolation) and model classification.
The full CT method's computation time only includes the inference time of the classification model.
Table~\ref{eff} reports the inference time for processing $50$ cases, where the percentage represents the proportion of the full CT method's time.
While Full CT yields high accuracy, its reliance on 3D convolutions results in prohibitive latency ($>169$s). 
CSS reduces this time by over $60\%$. 
Although slightly slower than simple heuristic methods (SI, SSS), CSS provides a far better trade-off, ensuring high diagnostic accuracy with acceptable latency, as shown in Table~\ref{others}.
Compared to ActiveFT, CSS is faster due to the efficient approximate nearest neighbor search.

\subsection{Visualization}

We visualize the CT slices selected by CSS in Fig.~\ref{vis}, where each case retains $4$ slices.
The selected slices accurately capture the lesion areas (marked in colored boxes) for all three diseases. 
Consistent with quantitative results, NCP and CP present larger, more obvious lesions, whereas AC manifests as smaller, localized nodules. 
The ability of CSS to pinpoint these subtle AC lesions confirms its effectiveness.

\section{Conclusions}

In this work, we introduce ERF, a framework designed to confront the dual impediments of computational expense and model trust that currently hinder the widespread clinical adoption of deep learning for pulmonary disease screening. Our approach makes two primary contributions. First, we propose CSS, which curates a compact yet diagnostically potent subset of CT slices by optimizing for both representativeness and diversity.
Second, we design a lightweight AUQ mechanism for our task, which captures data ambiguity arising from subtle lesions or artifacts.
Our extensive validation on two public datasets demonstrates that ERF not only matches the diagnostic performance of full-volume analysis but also reduces inference time by over $60\%$. More critically, ERF’s ability to reliably identify and flag uncertain predictions represents a pivotal step toward the deployment of fast, accurate, and trustworthy AI in real-world clinical workflows.

\bibliographystyle{IEEEbib}
\bibliography{icme2026references}

\section*{Appendix}

\subsection{Model Architecture of Coarse Screening}

First, we employ a ResNet-34~\cite{he2016deep} as the backbone for extracting features from each selected slice.
We modify the standard ResNet-34 by replacing the final $1000$-dimensional fully-connected (FC) layer with a $512$-dimensional FC layer and removing the final softmax function.
Then, the output of ResNet-34 is used as input to a light transformer-based network~\cite{vaswani2017attention} for feature fusion, where both the layer number and the head number of the Transformer Encoder are set to $6$.
Finally, we perform an FC layer and a softmax function on the output of the network to predict the probability of disease.

To ensure fair comparison, we configure the same classification model for all sub-sampling methods and compare the performance of models trained with data sub-sampled by different methods. Although the feature extraction process is at the slice level, we fuse them, and the model prediction output is at the patient level, so there is no data leakage problem, that is, slices from the same patient could not appear in both the training set and the validation set.
In addition, it is worth noting that the full CT method uses 3D-ResNet~\cite{fan2021pytorchvideo} as the classification model, therefore, there will be no data leakage problem.

\subsection{Implementations Details}

\subsubsection{Computer Resources}
All models are implemented in PyTorch and trained on an RTX 4090 with $24$ GB memory.

\subsubsection{Sub-Sampling}

\textbf{CSS:}
The CSS process is described in the main text, so we will not reiterate it here. Below, we provide the specific hyperparameter settings as follows:
$k=10$, $\alpha=0.5$, $\beta=0.9$, $\lambda=0.5$, and $T=10$.
It is worth noting that in the CSS method, the distance between each instance and its $k$-th nearest neighbor can be sensitive to noise, as it relies solely on the $k$-th nearest neighbor.
Following the settings of~\cite{wang2022unsupervised}, we modify this by using the average distance to the $k$ nearest neighbors, denoted as $\overline{D}(\bm{z}_i^j,k)=\frac{1}{k} \sum_{l=1}^k D(\bm{z}_i^j,\bm{z}_i^j|l)$.\\
\textbf{CoreSet~\cite{sener2018active} and ActiveFT~\cite{xie2023active}:}
For fair comparison, we set the preprocessing steps (\textit{i.e.}, feature extraction) and the lung patching process of the CoreSet and ActiveFT to be consistent with the CSS, and then perform slice selection according to their respective strategies. Their respective strategies can be found in the original paper, and are not described here again.\\
\textbf{PI, LI and SI:}
Interpolation methods calculate the new pixel value in the compressed image based on the values of surrounding pixels on the $Z$-axis.\\
\textbf{SSS:}
SSS selects $m/3$ slices from the first, middle, and last positions of the CT volume, respectively.\\
\textbf{ESS:}
ESS selects one after every $\lfloor n/m \rfloor$ of slices.

\subsubsection{Classification Model Training}
During the classification model training, we use Adam~\cite{adam2014method} with an initial learning rate $1e^{-3}$ to optimize the network.
We set different batch sizes for experiments with different numbers of selected slices. The more slices selected for each case, the more GPU memory each case occupies; thus, the batch size should be smaller. In detail, $\text{batch size} = 1,4,8,16,16$ is set for $m = 64, 32, 16, 8, 4$ respectively. For a fair comparison, the training epoch of all experiments is set to $400$.

\subsubsection{Uncertainty Quantification}

To ensure a fair comparison of the performance of UQ methods, the classification prediction outcomes of the models must remain consistent. Some UQ methods produce multiple outputs, such as the DE methods, which integrate predictions from multiple models, and the discrepancy-only methods, which employ multiple classifiers.
To address this, we establish a primary model (\textit{i.e.}, ResNet-34) for all methods, ensuring identical initialization and training configurations. Specifically, this model is trained using CT data sub-sampled via CSS, with the number of sub-sampled slices set to $32$. This approach guarantees consistent classification predictions across all methods while allowing different approaches to incorporate outputs from auxiliary branches during the UQ phase.\\
\textbf{Entropy-only method:}
The entropy of the predicted probability vector is used as the uncertainty measure. For sample $\widetilde{\bm{x}}_i$, entropy is calculated as:
$\mathcal{H}(\widetilde{\bm{x}}_i)=-\sum_{j=1}^{K}\bm{p}_{i,j}\log\bm{p}_{i,j}$,
where $\bm{p}_{i,j}$ denotes the $j$-th element of $\bm{p}_i$.\\
\textbf{Discrepancy-only method:}
This method employs the same model structure and training protocol as AUQ, but utilizes only discrepancy as the uncertainty metric. The discrepancy calculation is denoted as Equation (4) in the main text.\\
\textbf{DE method:}
Two auxiliary deep models are trained. Their architectures and training settings match the primary model, but with different initializations. During the UQ phase, the predictive probabilities of all three models are averaged, after which entropy is computed as the uncertainty measure.\\
\textbf{MC Dropout method:}
An additional neural network with Dropout layers is trained. It shares the same architecture as the primary model but includes Dropout layers ($\text{rate}=0.5$) after each block. During the UQ phase, this model performs two additional forward passes. Its outputs are averaged with those of the primary model, and entropy is subsequently computed as the uncertainty measure.

\begin{table*}[htbp]
\centering
\caption{Comparison under extreme sub-sampling.}
\resizebox{\textwidth}{!}{
\begin{tabular}{c|c|cc|cc|cc}
\toprule
\multicolumn{1}{c|}{}            & \multicolumn{1}{c|}{}
& \multicolumn{2}{c|}{Task $1$}    & \multicolumn{2}{c|}{Task $2$}   & \multicolumn{2}{c}{Task $3$}   \\
\multicolumn{1}{c|}{\multirow{-2}{*}{Number}}   & \multicolumn{1}{c|}{\multirow{-2}{*}{Method}}
& \multicolumn{1}{c}{Accuracy}   & \multicolumn{1}{c|}{Recall}
& \multicolumn{1}{c}{Accuracy}   & \multicolumn{1}{c|}{Recall}
& \multicolumn{1}{c}{Accuracy}   & \multicolumn{1}{c}{Recall}   \\
\cmidrule{1-8}
\multirow{3}{*}{$16$}
& SI
& \underline{$96.99_{\pm 1.55}$}           & \underline{$97.21_{\pm 1.10}$}
& \underline{$93.87_{\pm 2.27}$}           & $94.07_{\pm 2.03}$ 
& $76.71_{\pm 2.70}$           & $76.90_{\pm 2.56}$    \\
& ActiveFT
& $96.94_{\pm 0.87}$           & $97.19_{\pm 0.88}$ 
& $93.61_{\pm 1.48}$           & \underline{$94.20_{\pm 1.95}$} 
& \underline{$78.95_{\pm 2.11}$}           & \underline{$79.43_{\pm 2.03}$}    \\
& CSS (Ours)
& $\mathbf{97.26_{\pm 1.54}}$           & $\mathbf{97.58_{\pm 1.21}}$ 
& $\mathbf{94.00_{\pm 2.15}}$           & $\mathbf{94.28_{\pm 2.04}}$ 
& $\mathbf{85.60_{\pm 2.37}}$           & $\mathbf{85.54_{\pm 2.40}}$    \\
\cmidrule{1-8}
\multirow{3}{*}{$8$}
& SI
& $95.96_{\pm 1.19}$           & $95.88_{\pm 1.22}$ 
& $92.09_{\pm 2.77}$           & $92.55_{\pm 2.49}$ 
& $67.34_{\pm 3.83}$           & $67.68_{\pm 3.41}$    \\
& ActiveFT
& \underline{$96.13_{\pm 1.11}$}           & \underline{$96.12_{\pm 1.43}$} 
& \underline{$92.42_{\pm 1.75}$}           & \underline{$92.98_{\pm 2.28}$}
& \underline{$74.61_{\pm 3.13}$}           & \underline{$75.24_{\pm 2.41}$}    \\
& CSS (Ours)
& $\mathbf{96.57_{\pm 1.24}}$           & $\mathbf{96.89_{\pm 1.27}}$ 
& $\mathbf{93.21_{\pm 3.72}}$           & $\mathbf{93.02_{\pm 3.44}}$ 
& $\mathbf{82.55_{\pm 2.32}}$           & $\mathbf{82.63_{\pm 2.17}}$    \\
\cmidrule{1-8}
\multirow{3}{*}{$4$}
& SI
& $93.86_{\pm 2.24}$           & $94.24_{\pm 1.89}$ 
& $90.03_{\pm 2.25}$           & $90.78_{\pm 2.30}$ 
& $64.41_{\pm 3.70}$           & $64.57_{\pm 3.54}$    \\
& ActiveFT
& \underline{$94.07_{\pm 1.40}$}           & \underline{$94.51_{\pm 2.02}$} 
& \underline{$90.79_{\pm 2.08}$}           & \underline{$91.23_{\pm 2.44}$} 
& \underline{$70.72_{\pm 3.41}$}           & \underline{$69.14_{\pm 2.98}$}    \\
& CSS (Ours)
& $\mathbf{94.90_{\pm 2.12}}$           & $\mathbf{95.19_{\pm 2.01}}$ 
& $\mathbf{91.17_{\pm 2.54}}$           & $\mathbf{91.76_{\pm 2.51}}$ 
& $\mathbf{78.64_{\pm 3.19}}$           & $\mathbf{78.89_{\pm 2.98}}$    \\
\bottomrule
\end{tabular}
}
\label{cues}
\end{table*}

\subsection{Additional Comparison with Other Sub-Sampling Methods}

We compare CSS with SI and ActiveFT under extreme sub-sampling, that is, selecting $16$, $8$, and $4$ CT slices from each case.
The experimental results are shown in Table~\ref{cues}, from which we can observe that our method outperforms other methods, especially on the difficult task $3$.
For instance, at $\text{Number}=4$, CSS achieved an accuracy of $78.64\%$ and a recall of $78.89\%$.
Although this shows a significant drop compared to results with higher sampling rates, it surpasses the second-best method ActiveFT by $7.92\%$ in accuracy and $9.75\%$ in recall.
This demonstrates that under extreme sub-sampling scenarios, our method holds a greater advantage.

\end{document}